\documentclass[twocolumn,showpacs,aps,prl,superscriptaddress]{revtex4}

\usepackage{graphicx}
\usepackage{dcolumn}
\usepackage{amsmath}
\usepackage{epsfig}

\RequirePackage{xspace}

\usepackage{relsize}
\def\babar{\mbox{\slshape B\kern-0.1em{\smaller A}\kern-0.1em
    B\kern-0.1em{\smaller A\kern-0.2em R}}}

\def\epem       {\ensuremath{e^+e^-}\xspace}

\def\mumu       {\ensuremath{\mu^+\mu^-}\xspace}

\def\tautau     {\ensuremath{\tau^+\tau^-}\xspace}

\def\bbbar {\ensuremath{b\overline b}\xspace}

\def\pipi  {\ensuremath{\pi^+\pi^-}\xspace}

\def\Kbar  {\kern 0.2em\overline{\kern -0.2em K}{}\xspace}

\def\Kz    {\ensuremath{K^0}\xspace}
\def\Kzb   {\ensuremath{\Kbar^0}\xspace}
\def\KzKzb {\ensuremath{\Kz \kern -0.16em \Kzb}\xspace}
\def\Kp    {\ensuremath{K^+}\xspace}
\def\Km    {\ensuremath{K^-}\xspace}

\def\KpKm  {\ensuremath{\Kp \kern -0.16em \Km}\xspace}

\def\Dbar    {\kern 0.2em\overline{\kern -0.2em D}{}\xspace}

\def\Dz      {\ensuremath{D^0}\xspace}
\def\Dzb     {\ensuremath{\Dbar^0}\xspace}
\def\DzDzb   {\ensuremath{\Dz {\kern -0.16em \Dzb}}\xspace}
\def\Dp      {\ensuremath{D^+}\xspace}
\def\Dm      {\ensuremath{D^-}\xspace}

\def\DpDm    {\ensuremath{\Dp {\kern -0.16em \Dm}}\xspace}

\def\Bbar    {\kern 0.18em\overline{\kern -0.18em B}{}\xspace}

\def\Bz      {\ensuremath{B^0}\xspace}
\def\Bzb     {\ensuremath{\Bbar^0}\xspace}
\def\BzBzb   {\ensuremath{\Bz {\kern -0.16em \Bzb}}\xspace}
\def\Bu      {\ensuremath{B^+}\xspace}
\def\Bub     {\ensuremath{B^-}\xspace}

\def\BpBm    {\ensuremath{\Bu {\kern -0.16em \Bub}}\xspace}

\def\BorBbar    {\kern 0.18em\optbar{\kern -0.18em B}{}\xspace}
\def\DorDbar    {\kern 0.18em\optbar{\kern -0.18em D}{}\xspace}
\def\KorKbar    {\kern 0.18em\optbar{\kern -0.18em K}{}\xspace}

\mathchardef\Upsilon="7107
\def\Y#1S{\ensuremath{\Upsilon{(#1S)}}\xspace}
\def\OneS  {\Y1S}

\def\ThreeS{\Y3S}

\mathchardef\Deltares="7101
\mathchardef\Xi="7104
\mathchardef\Lambda="7103
\mathchardef\Sigma="7106
\mathchardef\Omega="710A

\def\Deltabar{\kern 0.25em\overline{\kern -0.25em \Deltares}{}\xspace}
\def\Lbar{\kern 0.2em\overline{\kern -0.2em\Lambda\kern 0.05em}\kern-0.05em{}\xspace}
\def\Sigbar{\kern 0.2em\overline{\kern -0.2em \Sigma}{}\xspace}
\def\Xibar{\kern 0.2em\overline{\kern -0.2em \Xi}{}\xspace}
\def\Obar{\kern 0.2em\overline{\kern -0.2em \Omega}{}\xspace}
\def\Nbar{\kern 0.2em\overline{\kern -0.2em N}{}\xspace}
\def\Xb{\kern 0.2em\overline{\kern -0.2em X}{}\xspace}

\newcommand{\tev}{\ensuremath{\mathrm{\,Te\kern -0.1em V}}\xspace}
\newcommand{\gev}{\ensuremath{\mathrm{\,Ge\kern -0.1em V}}\xspace}
\newcommand{\mev}{\ensuremath{\mathrm{\,Me\kern -0.1em V}}\xspace}
\newcommand{\kev}{\ensuremath{\mathrm{\,ke\kern -0.1em V}}\xspace}
\newcommand{\ev}{\ensuremath{\mathrm{\,e\kern -0.1em V}}\xspace}
\newcommand{\gevc}{\ensuremath{{\mathrm{\,Ge\kern -0.1em V\!/}c}}\xspace}
\newcommand{\mevc}{\ensuremath{{\mathrm{\,Me\kern -0.1em V\!/}c}}\xspace}
\newcommand{\gevcc}{\ensuremath{{\mathrm{\,Ge\kern -0.1em V\!/}c^2}}\xspace}
\newcommand{\mevcc}{\ensuremath{{\mathrm{\,Me\kern -0.1em V\!/}c^2}}\xspace}

\def\mus  {\ensuremath{\rm \,\mus}\xspace}

\def\mus        {\ensuremath{\,\mu{\rm s}}\xspace}    

\def\to                 {\ensuremath{\rightarrow}\xspace}

\def\pep2{PEP-II}

\def\gsim{{~\raise.15em\hbox{$>$}\kern-.85em
          \lower.35em\hbox{$\sim$}~}\xspace}
\def\lsim{{~\raise.15em\hbox{$<$}\kern-.85em
          \lower.35em\hbox{$\sim$}~}\xspace}

\xspace

\newcommand{\jprlBase}       {Phys.\ Rev.\ Lett.\xspace}

\newcommand{\jprl}      [1]  {\jprlBase\ {\bf #1}}

\def\evtgen     {\mbox{\tt EvtGen}\xspace}

\def\geant      {\mbox{\tt GEANT}\xspace}

\def\jetset74   {\mbox{\tt Jetset \hspace{-0.5em}7.\hspace{-0.2em}4}\xspace}

\newcommand{\BABARPubYear}    {09}
\newcommand{\BABARPubNumber}  {038}

\newcommand{\SLACPubNumber} {14008}
\newcommand{\LANLNumber} {1002.4358}

\def\figurebox#1#2#3{%
    \def\arg{#3}%
    \ifx\arg\empty
    {\hfill\vbox{\hsize#2\hrule\hbox to #2{\vrule\hfill\vbox to #1{\hsize#2\vfill}\vrule}\hrule}\hfill}%
    \else
    {\hfill\epsfbox{#3}\hfill}%
    \fi}

\begin{document}

\preprint{\babar-PUB-\BABARPubYear/\BABARPubNumber} 
\preprint{SLAC-PUB-\SLACPubNumber} 

\begin{flushleft}

\babar-PUB-\BABARPubYear/\BABARPubNumber\\
SLAC-PUB-\SLACPubNumber\\
arXiv:\LANLNumber\ [hep-ex]\\[10mm]
\end{flushleft}

\title{
{\large \bf  \boldmath
Test of lepton universality in $\OneS$ decays at \babar\ }
}

%
%
\author{P.~del~Amo~Sanchez}
\author{J.~P.~Lees}
\author{V.~Poireau}
\author{E.~Prencipe}
\author{V.~Tisserand}
\affiliation{Laboratoire d'Annecy-le-Vieux de Physique des Particules (LAPP), Universit\'e de Savoie, CNRS/IN2P3,  F-74941 Annecy-Le-Vieux, France}
\author{J.~Garra~Tico}
\author{E.~Grauges}
\affiliation{Universitat de Barcelona, Facultat de Fisica, Departament ECM, E-08028 Barcelona, Spain }
\author{M.~Martinelli$^{ab}$}
\author{A.~Palano$^{ab}$ }
\author{M.~Pappagallo$^{ab}$ }
\affiliation{INFN Sezione di Bari$^{a}$; Dipartimento di Fisica, Universit\`a di Bari$^{b}$, I-70126 Bari, Italy }
% ...
\collaboration{The \babar\ Collaboration}
\noaffiliation

\date{\today}

\begin{abstract}
The ratio $R_{\tau\mu}(\OneS) = \Gamma_{\OneS\to \tautau}/\Gamma_{\OneS\to \mumu}$ is measured using a sample of $(121.8\pm1.2)\times10^6$ $\ThreeS$ events recorded by the $\babar$ detector. This measurement is intended as a
test of lepton universality and as a search for a possible light pseudoscalar Higgs boson. 
In the standard model (SM) this ratio is expected to be close to 1. Any significant deviations would violate lepton universality and could be introduced by the coupling to a light pseudoscalar Higgs boson.
The analysis studies the decays $\ThreeS\to\OneS\pipi$, $\OneS\to l^+l^-$, where $l=\mu,\tau$.
The result, $R_{\tau\mu}(\OneS) = 1.005 \pm 0.013(stat.) \pm 0.022(syst.)$, shows no deviation from the expected SM value, while improving the precision with respect to previous measurements.
\end{abstract}


\maketitle

In the standard model (SM), the couplings of the gauge bosons to leptons are independent of the lepton flavor. 
Aside from small lepton-mass effects, the expression for the decay
width $\Upsilon(1S)\to l^+ l^-$ should be identical for all leptons, and given by~\cite{ref:mas}:
\begin{equation}
\Gamma_{\Upsilon(1S)\to ll}=4\alpha^2 Q_b^2 \frac{|R_n(0)|^2}{M^2_\Upsilon}(1+2\frac{M^2_l}{M^2_\Upsilon})\sqrt{1-4\frac{M^2_l}{M^2_\Upsilon}}, \label{eq:BR}
\end{equation}
where $\alpha$ is the electromagnetic fine structure constant, $Q_b$ is the charge of the bottom quark, $R_n(0)$ is the non-relativistic radial wave
function of the bound \bbbar\ state evaluated at the origin, $M_\Upsilon$ is the $\Upsilon(1S)$ mass and $M_l$ is the lepton mass. In the SM, one
expects the quantity
\begin{equation}
R_{ll'}(\Upsilon(1S)) =\frac{ \Gamma_{\Upsilon(1S)\to ll}}{\Gamma_{\Upsilon(1S)\to l'l'}}
\end{equation}
with $l,l'=e,\mu,\tau$ and $l'\neq l$, to be close to one. In particular, the value for $R_{\tau\mu}(\OneS)$ is predicted to be $\sim0.992$~\cite{ref:PDG2008}.

In the next-to-minimal extension of the SM~\cite{ref:Higgs}, deviations of $R_{ll'}$ from the SM expectation may arise due to a light CP-odd Higgs boson, $A^0$. Present data~\cite{ref:LEP} do not exclude the existence of such a boson with a mass below 10~$\gevcc$.
Among other hypothetical particles, $A^0$ may mediate the following processes~\cite{ref:mas}:
\begin{equation}
\OneS\to A^0\gamma\to l^+l^-\gamma\label{eqn:a0_nomix}
\end{equation}
or 
\begin{equation}
\OneS\to \eta_b(1S)\gamma, \eta_b(1S)\to A^0\to l^+l^-.\label{eqn:a0_etabmix}
\end{equation}
The latter implies a mixing between $A^0$ and $\eta_b(1S)$, which is a $^1S_0$ \bbbar\ state and therefore not expected to decay to a lepton pair to leading order in the SM.

If the photon is energetic enough to be detectable, a monochromatic peak in the photon spectrum recoiling against the lepton pair could be an indication of new physics (NP)~\cite{ref:A0mu,ref:A0tau}. Alternatively, if the photon remains undetected, the lepton pair would be ascribed to the $\OneS$ and the proportionality of the coupling of the Higgs to the lepton mass would lead to an apparent violation of  lepton universality. This effect should be larger for decays to $\tautau$ pairs, and enhanced for higher-mass $\Upsilon(nS)$ and $\eta_b(nS)$ resonances.
The deviation of $R_{ll'}$ from the expected SM value depends on $X_d=\cos\theta_A\tan\beta$ (where $\theta_A$ measures the coupling of the $\OneS$ to the $A^0$, and $\tan\beta$ is the ratio of the vacuum expectation values of the two Higgs doublets) and on the mass difference between $A^0$ and $\eta_b(1S)$. Assuming $X_d=12$ (a representative value evading present limits~\cite{ref:LEP}), $\Gamma(\eta_b(1S))=5$ MeV, and $M_{\eta_b(1S)}$ as measured in~\cite{ref:etab}, the deviation of $R_{\tau\mu}(\OneS)$ may be as large as $\sim4\%$, depending on the $A^0$ mass~\cite{ref:mas}.  A measurement of this ratio has already been performed, with the result $R_{\tau\mu}(\OneS) = 1.02 \pm 0.02 (stat.) \pm 0.05(syst.)$~\cite{ref:CLEO}.

This paper focuses on the measurement of $R_{\tau\mu}(\OneS)$, in the decays $\ThreeS\to\OneS\pipi$  with $\OneS\to l^+l^-$ and $l=\mu,\tau$.  In this analysis only $\tau$ decays to a single charged particle (plus neutrinos) are considered. This choice simplifies the analysis; in particular it results in final states of exactly four detected particles for both the $\mumu$ and $\tautau$ samples.
The data collected at the $\ThreeS$ resonance by the \babar\ detector at the \pep2 storage rings correspond to 28 fb$^{-1}$.
About one tenth of the complete available statistics is used to validate the analysis method and the signal extraction procedure.
This validation sample is discarded from the final result in order to avoid any possible bias. 
A sample of 2.4 fb$^{-1}$ collected about 30 MeV below the $\ThreeS$ resonance (off-resonance sample) is also used as a background control sample.

The \babar\ detector is described in detail elsewhere~\cite{ref:babar1, ref:babar2}.

The event selection is optimized using Monte Carlo (MC) simulated events, generated with \evtgen~\cite{ref:evtgen}. \geant~\cite{ref:geant} is used to reproduce interactions of particles traversing the \babar\ detector, taking into account the varying detector conditions and beam backgrounds.
Final state radiation (FSR) effects are simulated using \texttt{PHOTOS}~\cite{ref:photos}.

The selection requires exactly four charged tracks, each with transverse momentum $0.1<p_T<10~\gevc$, geometrically constrained to come from the same point. The distance of closest approach to the interaction region of each track must be less than 10 cm when projected along the beam axis and less than 1.5 cm in the transverse plane.
The ratio of the 2nd to 0th Fox-Wolfram moments (R2)~\cite{ref:foxwolfram} is required to be less than 0.97, and the absolute value of the cosine of the polar angle of the thrust axis~\cite{ref:thrust} to be less than 0.96.

A $\OneS\to l^+l^-$ candidate is formed by selecting two oppositely-charged tracks, constrained to come from a common vertex, and it is combined with two other oppositely-charged tracks, assigned the pion mass, to construct a $\ThreeS\to\OneS\pipi$ candidate.

Different selection criteria are used for the $\OneS\to\mumu$ and  the $\OneS\to\tautau$ decays, because in the latter the presence of neutrinos in the final state leads to a larger contamination from the background (mainly non-leptonic $\OneS$ decays  and $\epem\to\tautau$ events).
The $\OneS\to\mumu$ candidates ($D_\mu$) are selected by requiring two tracks in the final state identified as muons. This identification is performed by using information from different subdetectors, such as the energy deposited in the electromagnetic calorimeter, the number of hits 
in the instrumented flux return of the magnet and the number of interaction lengths traversed, combined in a neural-network algorithm. Calculated in the $\epem$ center-of-mass (CM) frame~\cite{ref:footnote1}, the difference between the initial state energy and the visible final state energy is required to be less than 0.5~\gev, the magnitude of the dipion momentum ($p_{\pi\pi}^*$)  less than $0.875~\gevc$, and the cosine of the angle between the two lepton candidates less than $-$0.96. 
For the $\OneS\to\tautau$ candidates ($D_\tau$), tighter selection criteria are applied to reduce background. In these events a large fraction of the energy is not reconstructed, due to the presence of neutrinos; thus the difference between the energy of the initial state and the energy detected in the final state, calculated in the $\epem$ CM frame, is required to exceed 5~\gev. Further requirements are made on the magnitude of the dipion momentum ($p_{\pi\pi}^*<0.825$~\gevc) and on the magnitude of the momentum of each $\pi$ ($p_{\pi}^*<0.725$~\gevc). The measured difference in the energy of the $\ThreeS$ and the $\OneS$ is restricted to $0.835 < \Delta E^* < 0.925$~\gev. 
A boosted decision tree~\cite{ref:BDT} is used to further reduce the background, based on several event shape and kinematic variables such as R2 and the energy of the charged tracks reconstructed in the events. The performance of the classifier is assessed using MC simulations and off-resonance data.

Finally, in order to select $\ThreeS\to\OneS\pipi$ candidates, the invariant mass difference $\Delta M = M(\ThreeS) - M(\OneS)$, calculated with the reconstructed tracks of the final state, is required to be less than 2.5~\gevcc and the dipion invariant mass ($M_{\pi\pi}$)  to be between 0.28 and 0.90~\gevcc.

For events with multiple candidates, the candidate with the value of $\Delta M$ closest to the nominal value~\cite{ref:PDG2008} is retained as the best one. 
It has been verified by MC simulations that the selection requirements do not reduce the sensitivity to NP processes.
Since the possible NP effects, with the presence of additional photons in the process, should be more evident in $\OneS\to\tautau$ events, variables that are sensitive to neutral energy are not used in the selection. 

The final selection efficiency for the reconstructed decay chains, estimated from a sample of MC simulated events, are $\epsilon_{\mu\mu}= (44.57\pm0.04)\%$  and $\epsilon_{\tau\tau}=(16.77\pm0.03)\%$ for the $\mumu$ and the $\tautau$ final states, respectively.

An extended unbinned maximum likelihood fit, applied simultaneously to the two disjoint datasets $D_\mu$ and $D_\tau$, is used to extract  $R_{\tau\mu}= \frac{N_{sig\tau}}{\epsilon_{\tau\tau}}\cdot\frac{\epsilon_{\mu\mu}}{N_{sig\mu}}$, where $N_{sig\mu}$ ($N_{sig\tau}$) indicates the number of signal events in the $D_\mu$ ($D_\tau$) sample. 
For the $D_\mu$ sample, a 2-dimensional probability density function (PDF) is used, based on the invariant dimuon mass $M_{\mumu}$ and $M_{\pipi}^{reco}$, the invariant mass of the system recoiling against the pion pair, defined as:
\begin{equation}
M_{\pipi}^{reco} = \sqrt{s+M_{\pi\pi}^2-2\cdot\sqrt{s}\cdot E_{\pi\pi}^*}, \label{eqn:MReco}
\end{equation}
where $\sqrt{s}$ is the $\epem$ center-of-mass energy and $E_{\pi\pi}^*$ indicates the $\pipi$ pair energy.
MC simulations are used to check that the two variables are uncorrelated.
For  the $D_\tau$ sample,  a 1-dimensional PDF is used, based on $M_{\pipi}^{reco}$ (Eq.~\ref{eqn:MReco}).
The likelihood is written as:
\begin{equation}
{\cal L}_{ext}  =  {\cal L}_{ext}^{\mu}\cdot {\cal L}_{ext}^{\tau},
\label{eq:likelihood}
\end{equation}
where:
\begin{equation}
{\cal L}_{ext}^{i} = \frac{e^{-N'_i}(N'_i)^{N_i}}{N_i!} \prod_{k=1}^{N_i} {\cal P}^i_k,
\label{eq:likelihood2}
\end{equation}
with $i=\mu$ or $\tau$ and where $N_i$ and $N'_i$ are the sum of the signal and background events, observed and expected respectively, in each sample. 
${\cal P}_k$ is the probability to measure a set of physical observables in the $k^{th}$ event, defined as:
\begin{eqnarray}
{\cal P}_k^\mu & \equiv & \frac{N_{sig\mu}}{N'_{\mu}} {\cal P}_k^{\mu}(M_{\pi^+\pi^-}^{reco})\cdot {\cal P}_k^{\mu}(M_{\mu^+\mu^-}) + \nonumber \\ 
 & + & \frac{N_{bkg\mu}}{N'_{\mu}} {\cal P}_k^{bkg\mu}(M_{\pi^+\pi^-}^{reco}) \cdot {\cal P}_k^{bkg\mu}( M_{\mu^+\mu^-}) \label{eqn:LHmu}
\end{eqnarray}
and: 
\begin{eqnarray}
{\cal P}_k^\tau & \equiv  & \frac{\epsilon_{\tau\tau}}{\epsilon_{\mu\mu}}\frac{N_{sig\mu}}{N'_{\tau}}R_{\tau\mu}{\cal P}_k^{\tau}(M_{\pi^+\pi^-}^{reco}) +\nonumber \\
 & + & \frac{N_{bkg\tau}}{N'_{\tau}}{\cal P}_k^{bkg\tau}(M_{\pi^+\pi^-}^{reco})  \label{eqn:LHtau}
\end{eqnarray}
where $N_{bkg\mu}$ ($N_{bkg\tau}$) indicates the number of background events in the $D_\mu$ ($D_\tau$) sample.

The functional forms of the PDFs describing the signal components are modeled from the dedicated sub-sample consisting of one tenth of the $D_\mu$ sample. 
Both the $M_{\pi^+\pi^-}^{reco}$ and the $M_{\mumu}$ distributions are described by an analytical function approximating a Gaussian distribution function with mean value $\mu$ but different left and right widths, $\sigma_{L,R}$, plus asymmetric non-Gaussian tails $\alpha_{L,R}$, defined as:
\begin{equation}
{\cal F}(x) = exp\Big\{ -\frac{(x-\mu)^2}{2\sigma^2_{L,R}+\alpha_{L,R}(x-\mu)^2} \Big\}. \label{eqn:Cruijff}
\end{equation}
All the parameters (the five parameters describing the $M_{\mumu}$ distribution, along with the mean values and the widths of both the $M_{\pi^+\pi^-}^{reco}$ distributions) are free in the fit, except for $\alpha_{L,R}$ in $M_{\pi^+\pi^-}^{reco}$.
The off-resonance sample is used to model the background shapes. Constants are chosen for the $D_\mu$ sample, and a first order polynomial for the $D_\tau$ sample, with all the parameters free in the fit. 

\begin{figure}[!htb]
\begin{center}
\includegraphics[width=0.9\linewidth]{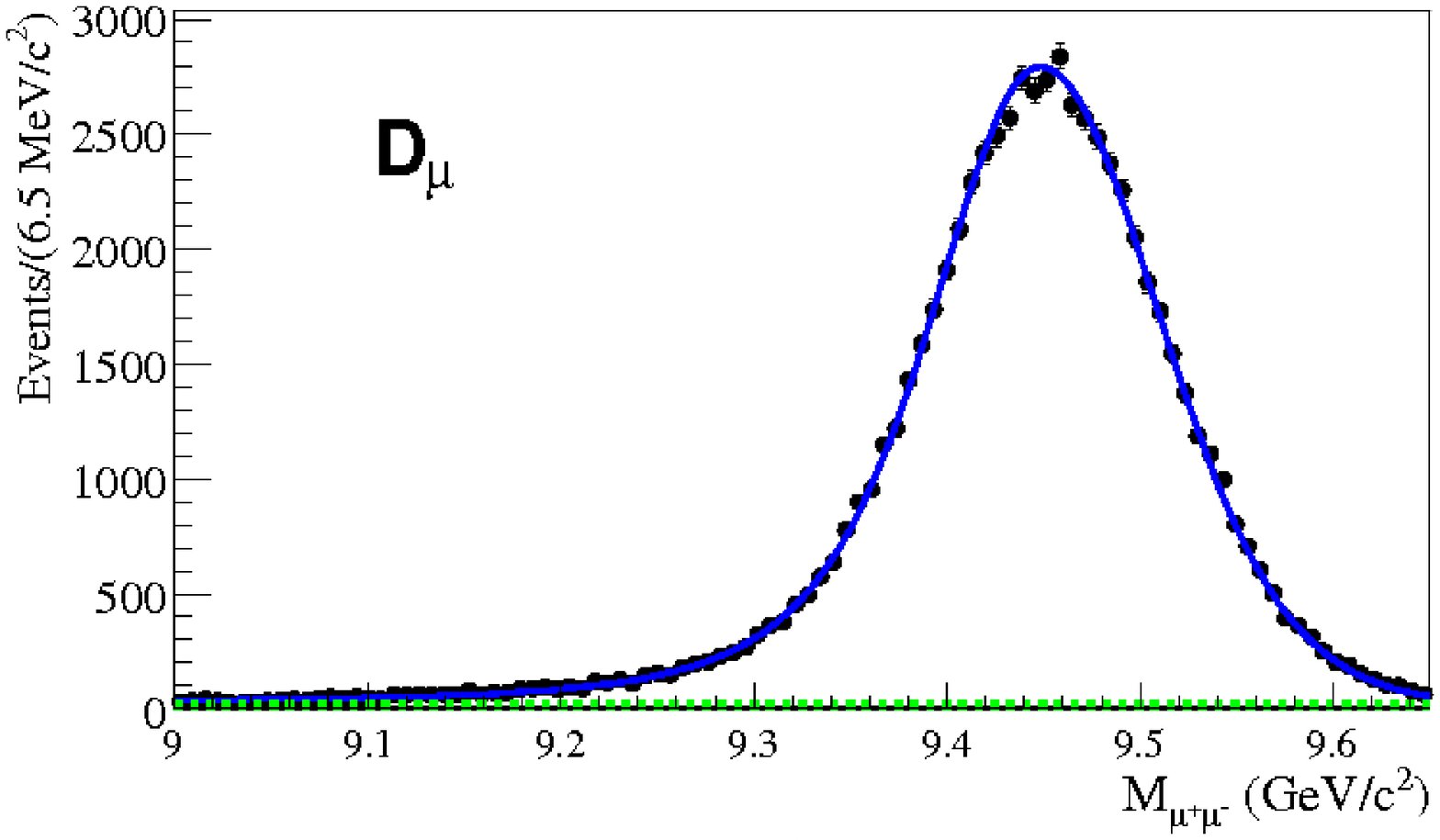}\\
\includegraphics[width=0.9\linewidth]{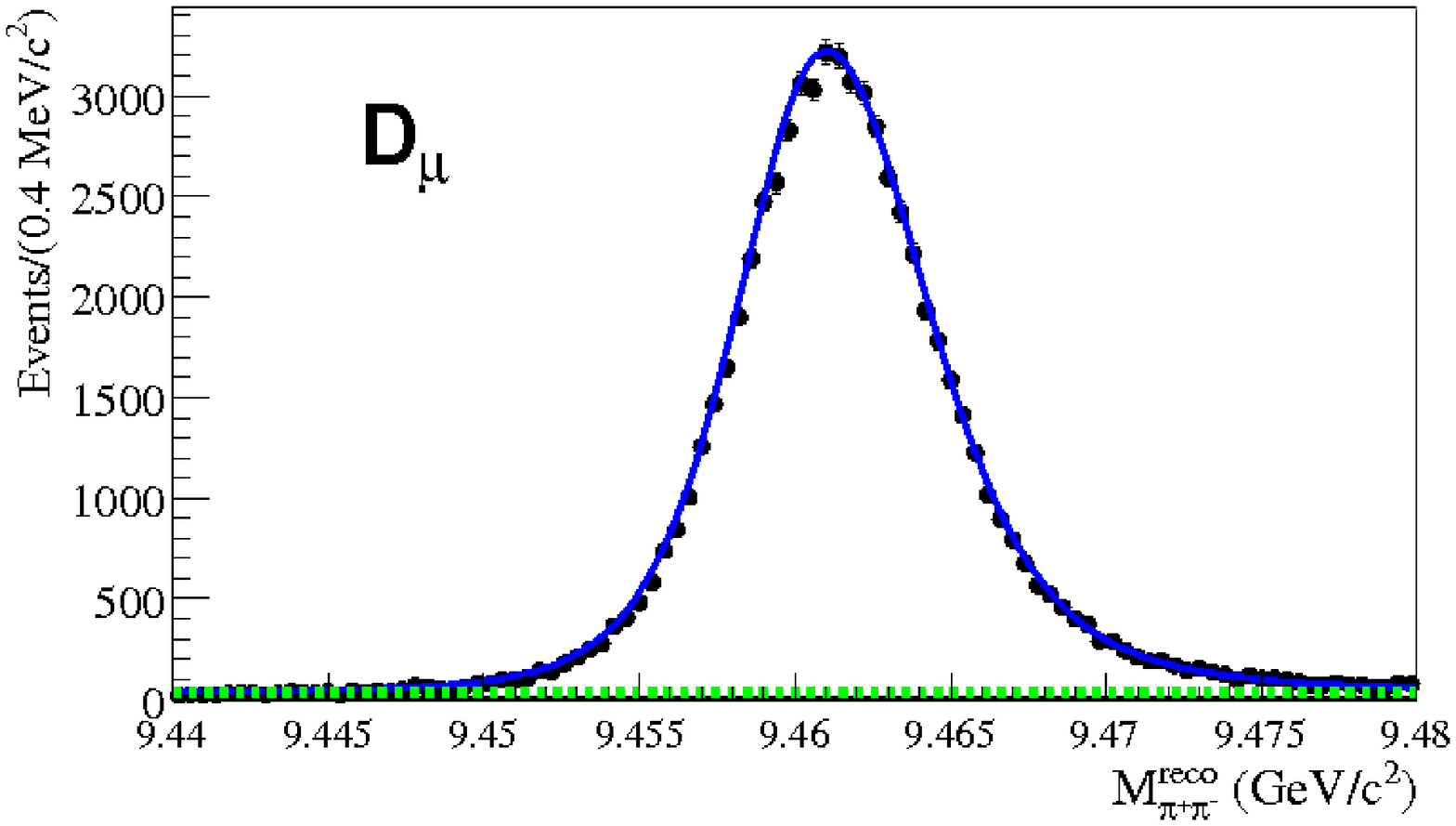}\\
\includegraphics[width=0.9\linewidth]{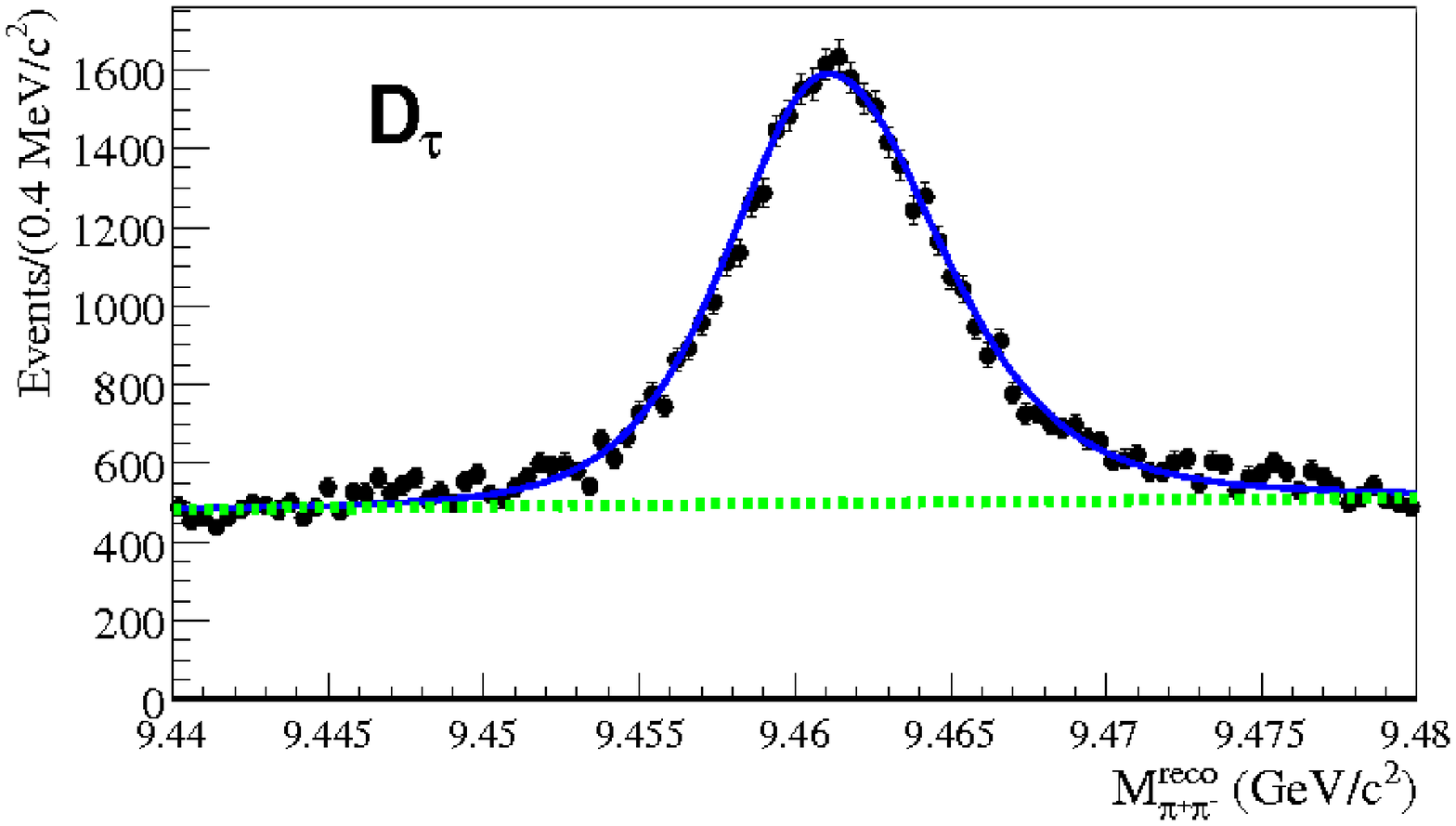}\\
\caption{1-D fit projections for $M_{\mumu}$ (top) and for $M_{\pipi}^{reco}$ (middle) in the $D_\mu$ sample, and for $M_{\pipi}^{reco}$ (bottom) in the  $D_\tau$ sample. In each plot the dashed  line represents the background shape, while the solid line is the sum of signal and background contributions to the fit, and the points are the data. 
}
\label{fig:fit}
\end{center}
\end{figure}

The result of the simultaneous fit is $R_{\tau\mu}=1.006\pm0.013$, where the quoted error is statistical only.
Figure~\ref{fig:fit} shows the projections of the fit results for the three variables.

%

Several systematic errors cancel in the ratio, such as errors on the luminosity, the $\ThreeS$ production cross section, and the $\ThreeS\to\OneS\pipi$ branching fractions, as well as systematic discrepancies between data and simulation in the common event selection and in track reconstruction efficiencies, where a possible dependence on the track energy has been taken into account.
The residual systematic uncertainties are related to the differences between data and simulation in the efficiency of event selection, 
the muon identification, and the trigger and background digital filters (BGF)~\cite{ref:footnote2}. There is also a systematic uncertainty on the signal and background yields due to the imperfect knowledge of the PDFs used in the fit.

The systematic uncertainty due to the event selection is evaluated by comparing the shape of each variable between data and simulation and estimating the difference in the efficiency.
The resulting systematic uncertainty is $1.2\%$.

The systematic uncertainty related to the difference between data and simulation of muon identification efficiencies, applied only to $\OneS\to\mumu$ events, is estimated by using two independent samples: one where both leptons are required to be identified as muons, and another where exactly one final charged track is a muon. The ratio of the efficiencies for requiring each sample is determined, both on data and on simulation; the ratio of the two results gives an efficiency correction of 1.023 and a related systematic uncertainty of  $1.2\%$.

The systematic uncertainties due to the differences between data and simulation in trigger and BGF efficiency are small both in $\OneS\to\mumu$ and in $\OneS\to\tautau$ events, and they cancel partially in the ratio. 
A correction of 1.020 is needed for the $\epsilon_{\tau\tau}$ efficiency, together with a systematic uncertainty of $0.10\%$ for $\OneS\to\tautau$ events, while a systematic uncertainty of $0.18\%$ is quoted for $\OneS\to\mumu$ events. The impact of the uncertainty in the BGF efficiency has been found to be negligible.

The uncertainty due to the imperfect knowledge of the signal and background shapes used in the fit is also estimated. The systematic effect from fixing $\alpha_{L,R}$ in the signal $M_{\pipi}^{reco}$ PDF is estimated by varying the fixed parameter values by $\pm 1\sigma$ and repeating the fit procedure. Since the correlation between the parameters is found to be negligible, the parameters are varied independently and the deviations from the nominal fit are summed in quadrature, resulting in a total  effect of  $1.1\%$.
The uncertainty due to the choice of the background PDF shapes is evaluated to be $0.22\%$, by using alternative parameterizations.
In the fit, the same $M_{\pipi}^{reco}$ functional form is used for both the $D_\mu$ and the $D_\tau$ sample, ignoring the potential difference in the  trigger efficiency. The systematic uncertainty associated with this approximation is evaluated to be $0.6\%$, by re-weighting the parameters for the $M_{\pipi}^{reco}$ distribution with the  parameters obtained from the $\tautau$ data sample, and requiring the magnitude of the momentum of one of the final state charged tracks not to exceed $1\gevc$.

The $M_{\pipi}^{reco}$ variable is related only to the $\ThreeS\to\OneS\pipi$ transition and therefore cannot distinguish between $\OneS\to l^+l^-$ events and other $\OneS$ decays or the Higgs-mediated events of Eq.~\ref{eqn:a0_nomix} and Eq.~\ref{eqn:a0_etabmix}. 
While this ensures sensitivity to possible NP effects, $\OneS$ generic decays could be a relevant source of background in the $D_\tau$ sample because the final state is only partially reconstructed.
The event selection heavily reduces the yield of  the $\OneS$ generic decays. It is estimated using a simulated sample of inclusive $\OneS$ decays, and is found to be $\sim0.4\%$ of the $\OneS\to\tau^+\tau^-$ signal yield. Since the hadronic $\OneS$ decays are not well measured, the simulation may not be reliable and a systematic uncertainty needs to be considered. A correction factor of 0.996, taking into account this contribution, is applied to the $\OneS\to\tau^+\tau^-$ signal yield, and a systematic uncertainty equal to $0.4\%$ is included as well.

The systematic uncertainty associated with the simulation of the FSR by \texttt{PHOTOS} is found to be negligible.
 
Finally, the finite size of the MC-simulated samples used to determine the efficiencies gives a contribution to the systematic uncertainty less than $0.1\%$ in both the leptonic final states.

The total systematic uncertainty, obtained by summing in quadrature all the contributions, is estimated to be $2.2\%$.

Including all the systematic corrections, the ratio $R_{\tau\mu}$ is found to be:
\[
R_{\tau\mu}(\OneS)=1.005 \pm 0.013(stat.) \pm 0.022(syst.).
\]
No significant deviation of the ratio $R_{\tau\mu}$ from the SM expectation is observed.
This result improves both the statistical and systematic precision with respect to the previous measurement~\cite{ref:CLEO}.
According to~\cite{ref:mas}, and assuming values for  $X_d$, $\Gamma(\eta_b(1S))$ and $M_{\eta_b(1S)}$ as previously stated, the present measurement excludes an $A^0$ with mass lower than 9\gevcc at 90$\%$ of confidence level.

We are grateful for the excellent luminosity and machine conditions
provided by our \pep2\ colleagues, 
and for the substantial dedicated effort from
the computing organizations that support \babar.
The collaborating institutions wish to thank 
SLAC for its support and kind hospitality. 
This work is supported by
DOE
and NSF (USA),
NSERC (Canada),
CEA and
CNRS-IN2P3
(France),
BMBF and DFG
(Germany),
INFN (Italy),
FOM (The Netherlands),
NFR (Norway),
MES (Russia),
MEC (Spain), and
STFC (United Kingdom). 
Individuals have received support from the
Marie Curie EIF (European Union) and
the A.~P.~Sloan Foundation.


\begin{thebibliography}{99}

\bibitem{ref:mas} M.~A.~Sanchis-Lozano, Int. J. Mod. Phys. {\bf A19}, 2183  (2004);  E.~Fullana and M.~A.~Sanchis-Lozano, Phys.\ Lett.\  {\bf  B653}, 67 (2007); F.~Domingo {\it et al.}, JHEP {\bf 0901}, 061 (2009).
\bibitem{ref:PDG2008}  C.~Amsler {\it et al.}  (Particle Data Group),  Phys.\ Lett.\  {\bf B667}, 1 (2008).
\bibitem{ref:Higgs} R.~Dermisek and J.~F.~Gunion,  \jprl{95}, 041801 (2005).  
\bibitem{ref:LEP} S.~Kraml {\it et al.}, CERN 2006-009; S.~Schael  {\it et al.} (ALEPH and DELPHI Collaborations), Eur. Phys. J. {\bf C47}, 547 (2006).
\bibitem{ref:A0mu} B.~Aubert {\it et al.}  (\babar\ Collaboration), \jprl{103}, 081803 (2009).
\bibitem{ref:A0tau} B.~Aubert {\it et al.}  (\babar\ Collaboration), \jprl{103}, 181801 (2009).
\bibitem{ref:etab}  B.~Aubert {\it et al.}  (\babar\ Collaboration), \jprl{101}, 071801 (2008).
\bibitem{ref:CLEO}  D.~Besson {\it et al.} (CLEO Collaboration), \jprl{98}, 052002 (2007).
\bibitem{ref:babar1}  B.\ Aubert {\em et al.} (\babar\ Collaboration), Nucl.\ Instrum.\ Methods {\bf A479}, 1 (2002).
\bibitem{ref:babar2} W.\ Menges,  IEEE Nucl.\ Sci.\ Symp.\ Conf.\ Rec.\  {\bf 5}, 1470 (2006).
\bibitem{ref:evtgen}
  D.~J.~Lange,  Nucl.\ Instrum.\ Meth.\  A {\bf 462}, 152 (2001).
\bibitem{ref:geant}  S.~Agostinelli {\it et al.}  (GEANT4 Collaboration), Nucl.\ Instrum.\ Meth.\  {\bf A506}, 250 (2003).
\bibitem{ref:photos}  E.~Barberio and Z.~Was, Comput.\ Phys.\ Commun.\  {\bf 79}, 291 (1994).
\bibitem{ref:foxwolfram}   G.~C.~Fox and S.~Wolfram, Nucl. Phys. \textbf{B149}, 413 (1979) [Erratum-ibid.\  {\bf B157}, 543 (1979)].
\bibitem{ref:thrust}   S.~Brandt {\it et al.}, Phys. Lett. \textbf{12}, 57 (1964).
%
\bibitem{ref:footnote1} An asterisk ($*$) is used to denote variables calculated in the $\epem$ CM frame.
\bibitem{ref:BDT} H.~J.~Yang, B.~P.~Roe and J.~Zhu, Nucl.\ Instrum.\ Meth.\  {\bf A574}, 342 (2007).
%
\bibitem{ref:footnote2} The background filters are applied to the data in order to reduce the contribution of specific backgrounds. They consist of several requirements on the number, energy, and momentum of the charged tracks and neutral clusters in the event.



\end{thebibliography}
\end{document}